\documentstyle[11 pt]{article}
\newcommand{\be}{\begin{equation}}
\newcommand{\ee}{\end{equation}}
\newcommand{\bea}{\begin{eqnarray}}
\newcommand{\eea}{\end{eqnarray}}
\newcommand{\sptwo}{1.4}
\newcommand{\doublespace}{\edef\baselinestretch{\sptwo}\Large\normalsize}
\begin{document}
\vspace{1.0in}

\begin{center}
{\bf Nonlinear Realizations of Supersymmetry and Other Symmetries}
\end{center}
\begin{center} 

S.T. Love\footnote{e-mail address: loves@physics.purdue.edu} \\
{\it Department of Physics\\ 
Purdue University\\
West Lafayette, IN 47907-2306}
~\\
\end{center}
\vspace{.3in}
\begin{abstract}
\noindent
Simultaneous nonlinear realizations of spontaneously broken supersymmetry in conjunction with other spontaneous and/or explicitly broken symmetries including $R$ symmetry, global chiral symmetry, dilatations and the superconformal symmetries is reviewed.

\end{abstract}

\vspace{.3in}

\doublespace

If supersymmetry (SUSY) is to be realized in nature, then it must be as a broken symmetry. Moreover, in order to be able to consistently gauge the symmetry thus producing a supergravity, the breaking must be spontaneous. Consequently, its dynamics necessarily includes a Nambu-Goldstone fermion, the Goldstino. An extremely useful, model independent, way of ensapsulating the dynamical constraints mandated by the spontaneous symmetry breakdown is through the use of effective Lagrangians based on nonlinear realizations\cite{SW}-\cite{CCWZ}.  

The supersymmetry (graded) algebra
\bea
&&\{ Q_\alpha , \bar{Q}_{\dot\alpha} \} = 2\sigma^\mu_{\alpha \dot\alpha} P_\mu ~~;~~\{ Q_\alpha , Q_\beta \}=\{ \bar{Q}_{\dot\alpha} , \bar{Q}_{\dot\beta} \}=0 \cr
&&[P^\mu, Q_\alpha]=[P^\mu, \bar{Q}_{\dot\alpha}]=[P^\mu, R]=0 \cr
&&[R, Q_{\alpha} ] = Q_{\alpha}~~;~~[R, \bar{Q}_{\dot\alpha} ] = -\bar{Q}_{\dot\alpha} \, ,
\label{SSA}
\eea
includes the unbroken space-time generators $P^\mu$ along with both the SUSY charges $Q_\alpha ,~\bar{Q}_{\dot\alpha}$ and the $R$ charge. 

Consider some unspecified underlying theory in which both the supersymmetry (SUSY) and the $R$ symmetry are spontaneously broken. This allows for the possibility that the dynamics producing the SUSY breaking has a different origin than that producing the $R$ symmetry breaking and moreover that the scales at which the symmetry breakings occur could be independent. The low energy degrees of freedom comprising the relevant effective Lagrangian include the Goldstino, the Nambu-Goldstone fermion of spontaneously broken supersymmetry\cite{AV}-\cite{FI} described by the two-component Weyl spinor fields $\lambda^\alpha, \bar\lambda_{\dot\alpha}$ and the $R$-axion, $a$, the (pseudo-) Nambu-Goldstone boson of spontaneously and soft explicitly broken $R$ symmetry. If the spontaneously broken supersymmetry is gauged to form a supergravity, the erstwhile Goldstino degrees of freedom  are absorbed to become the longitudinal (spin 1/2) modes of the spin 3/2 gravitino via the super-Higgs mechanism. As such the dynamics of those modes are given by that of the Goldstino. In fact, it is through the Goldstino degrees of freedom that SUSY might be most readily revealed experimentally\cite{phenom}. This brief review focuses on the nonlinear realization of spontaneously broken $N=1$ SUSY\cite{extsusy} in four flat space-time dimensions\cite{curvedsusy} in the presence of other spontaneously and/or explicitly broken symmetries. Included in this class are $R$ symmetry, superconformal symmetry, global chiral symmetries and dilatations. 

One method\cite{CCWZ} for constructing nonlinear realizations of spontaneously broken internal symmetries employs the construction of the coset group element parametrized by the coset space coordinates which are the associated Nambu-Goldstone bosons and then extracting the changes in these coordinates under group multiplication. Here this very general procedure is applied to construct nonlinear realizations of SUSY and $R$ symmetry\cite{AV},\cite{CL1},\cite{WS}. Since these are spontaneously broken spacetime symmetries, the motion in the coset space, whose coordinates are the $R$ axion and the Goldstinos, is accompanied by motion in spacetime. Thus it is useful to include the product of the unbroken translation group element along with the coset group elements and define
\be
\Omega(x,\lambda,\bar\lambda, a) = e^{-ix_\mu P^\mu} e^{\frac{i}{f_s^2}(\lambda^\alpha(x) Q_\alpha + \bar\lambda_{\dot\alpha}(x) \bar{Q}^{\dot\alpha}) }e^{\frac{i}{f_a}a(x) R} 
\, .
\ee
where $f_s$ and $f_a$ are the SUSY and $R$-symmetry breaking scales respectively. 
To extract the motion in the coset space, consider the product $g\Omega$ where 
$g= e^{i(\xi^\alpha Q_\alpha +\bar\xi_{\dot\alpha} \bar{Q}^{\dot\alpha})} e^{i\rho R}$ is a group element parametrized by the  space-time independent 2-component Weyl spinors $\xi^\alpha, \bar\xi_{\dot\alpha}$ for the SUSY transformations and $\rho$ for the $R$-transformation. This product of group elements again takes the form of a product of a translation and a coset group element but with translated spacetime points and coset coordinates so that 
\be
 g\Omega(x, \lambda, \bar\lambda, a) =  e^{-ix_{ \mu}^\prime P^\mu} e^{\frac{i}{f_s^2}(\lambda^{\prime\alpha}(x^\prime) Q_\alpha + \bar\lambda^\prime_{\dot\alpha}(x^\prime)  \bar{Q}^{\dot\alpha} )}e^{ \frac{i}{f_a}a^\prime(x^\prime) R}=\Omega (x^\prime, \lambda^\prime, \bar\lambda^\prime, a^\prime) .
\label{gt}
\ee
The total ($\Delta$) variation  of a generic field, $\phi^i(x)$, is defined as \\
$\Delta\phi^i(x)=\phi^{i\prime}(x^\prime)-\phi^i(x)$. Under nonlinear SUSY, one extracts
\be
\Delta_Q(\xi,\bar\xi) \lambda^\alpha (x)= f_s^2\xi^\alpha ;~ \Delta_Q(\xi,\bar\xi) \bar\lambda_{\dot\alpha} (x) = f_s^2\bar\xi_{\dot\alpha} ;~\Delta_Q(\xi,\bar\xi) a(x)=0 ~,
\ee
 while under $R$ transformations, 
\be
\Delta_R(\rho)\lambda^\alpha (x)= i\rho\lambda^\alpha (x);~\Delta_R(\rho)\bar\lambda_{\dot\alpha}(x) = -i\rho\bar\lambda_{\dot\alpha}(x); ~\Delta_R(\rho) a(x) = f_a\rho ~.
\ee
 The inhomogeneous terms in the various transformation are characteristic of the Nambu-Goldstone realization of the associated symmetries. The accompanying movement in spacetime, $\Delta x^\mu =x^{\prime\mu}-x^\mu$, is given by   
\be
\Delta_Q(\xi,\bar\xi) x^{\mu} = \frac{i}{f_s^2}[\lambda (x) \sigma^\mu \bar\xi -\xi \sigma^\mu \bar\lambda (x)]\equiv -\Lambda^\mu(\xi,\bar\xi)~~;~~\Delta_R(\rho) x^\mu =0 ~.
\label{SSV}
\ee
It also proves useful to define the intrinsic ($\delta$) variations of the fields as 
\be
\delta \phi^i (x) = \phi^{i\prime}(x)-\phi^i(x)=\Delta \phi^i (x) - \Delta x^\mu \partial_\mu \phi^i (x) ~.
\label{IV}
\ee
Any field, $\phi^i$, whose intrinsic variation under nonlinear SUSY is given by $\delta_Q(\xi,\bar\xi) \phi^i (x)=\Lambda^\mu(\xi,\bar\xi)\partial_\mu
\phi^i $  is said to carry the standard realization. 

From Eq. ($\ref{gt}$), it follows that the algebra-valued Maurer-Cartan 1-form, $\Omega^{-1}d\Omega$, is a total ($\Delta$) invariant under translations, SUSY and $R$ transformations: $(\Omega^{-1}d\Omega)^\prime(x^\prime)=(\Omega^{-1}d\Omega)(x)$. Expanding in terms of the translation, SUSY and $R$ generators as
\be
\Omega^{-1}d\Omega =i [-\omega_P^\mu (x) P_\mu +\frac{1}{f_s^2}\omega_Q^\alpha (x)Q_\alpha + \frac{1}{f_s^2}\bar\omega_{\bar{Q}~\dot\alpha}(x)  \bar{Q}^{\dot\alpha} +\frac{1}{f_a}\omega_R(x) R] \, ,
\ee
and employing the Baker-Campbell-Hausdorff formula, the coefficient one-forms are extracted as
\bea
\omega_P^\nu (x)&=& dx^\mu [\eta_\mu^\nu + \frac{i}{f_s^4}\lambda\stackrel{\leftrightarrow}{\partial}_\mu\sigma^\nu\bar{\lambda}]\equiv dx^\mu A_\mu~^\nu (x) \cr
\omega_Q^\alpha (x) &=& dx^\mu e^{-\frac{i}{f_a}a(x)}  \partial_\mu \lambda^\alpha (x) = \omega_P^\mu(x) e^{-\frac{i}{f_a}a(x)}  D_\mu\lambda^\alpha(x) =\omega_P^\mu(x)  \nabla_\mu\lambda^\alpha(x)\cr
\bar\omega_{\bar{Q}~\dot\alpha}(x)&=& e^{\frac{i}{f_a}a(x)}  \partial_\mu \bar\lambda_{\dot\alpha}(x) =\omega_P^\mu(x)e^{\frac{i}{f_a}a(x)} D_\mu\bar\lambda_{\dot\alpha}(x)=\omega_P^\mu(x) \nabla_\mu\bar\lambda_{\dot\alpha}(x)\cr
\omega_R (x) &=& dx^\mu  \partial_\mu a(x)=\omega_P^\mu (x) D_\mu a(x)
\, .
\eea
As a consequence of the $\Delta$ invariance of the Maurer-Cartan form, $\omega_P^\mu$ and the covariant derivatives $\nabla_\mu\lambda, \nabla_\mu\bar\lambda, D_\mu a$ are all invariant under the total $\Delta$-variations, while the covariant derivatives transform as the standard realization under the intrinsic nonlinear SUSY  variations while being $R$ invariant.  

Using the $\Delta$ invariance of $\omega_P^\mu$ along with $dx^{\prime ~\mu} =\frac{\partial x^{\prime~\mu}}{\partial x^\nu}dx^\nu = dx^\nu (\eta_\nu^\mu -\partial_\nu \Lambda^\mu)\equiv dx^\nu G_{\nu}~^\mu $, it follows that $A^\prime_\mu ~^\nu (x^\prime) = G^{-1}_\mu~^\rho (x)A_\rho~^\nu (x)$ and   
thus the product $d^4x \det{A}$ is $\Delta$ invariant: $d^4x^\prime \det{A^\prime}(x^\prime)=d^4x \det{A}(x)$. Consequently,  $A_\mu~^\nu = \eta_\mu^\nu + \frac{i}{f_s^4}\lambda\stackrel{\leftrightarrow}{\partial}_\mu\sigma^\nu\bar{\lambda}$ can be viewed as a ``vierbein'' and an  action invariant under both nonlinear SUSY and $R$-transformations can be constructed as
\be
\Gamma =\int d^4x {\cal L}=\int d^4x (\det{A})~{\cal O}(Da,\nabla\lambda,\nabla\bar\lambda)
\ee
with ${\cal O}$ is any Lorentz singlet function. The leading terms in a derivative expansion of the Lagrangian are
\be
{\cal L} = -\frac{f_s^4}{2}\det{A}~-\frac{1}{2}\det{A}~D_\mu a D^\mu a \, .
\label{AV}
\ee
In the absence of the $R$ axion, Eq. (\ref{AV}) reduces to the Akulov-Volkov action\cite{AV}. A nonlinearly SUSY invariant but soft $R$-symmetry breaking mass term can also be included producing the Lagrangian 
\be
{\cal L} = -\frac{f_s^4}{2}\det{A}~-\frac{1}{2}\det{A}~D_\mu a D^\mu a -\frac{1}{2}m_a^2 \det{A}~ a^2 \ .
\label{L1}
\ee

For any  phenomenological application, it is necessary to determine the coupling the Nambu-Goldstone modes to generic matter fields ($\psi^i$)\cite{CL2} and to gauge fields $B_\mu^a$\cite{CLLW}-\cite{BZF}. The matter fields transform as a standard realization under nonlinear SUSY and carry R weight $n_\psi$. The nonlinear SUSY covariant derivative for the matter fields is then defined as $(\nabla_\mu \psi)^i = e^{-\frac{i}{f_a}n_\psi a}D_\mu\psi^i = e^{-\frac{i}{f_a}n_\psi a }A^{-1}_\mu~^\nu \partial_\nu \psi^i $, so that it also transforms as the standard realization under nonlinear SUSY while being $R$ invariant. A combined nonlinear SUSY and gauge covariant derivative can then be introduced as \\
$({\cal D}_\mu \psi)^i = e^{-in_\psi \frac{a}{f_a}}(A^{-1})_\mu~^\nu [\partial_\mu\psi^i +{T}^a_{ij}{B}^a_\mu \psi^j ]$. So doing, $({\cal D}_\mu \psi)^i$ also transforms as the nonlinear SUSY standard realization provided the vector potential has the SUSY transformation $\delta_Q(\xi,\bar{\xi})B^a_\mu = 
\Lambda^\rho (\xi,\bar\xi)\partial_\rho  B^a_\mu +\partial_\mu \Lambda^\rho (\xi,\bar\xi)
B^a_\rho $. Alternatively, one can introduce the redefined gauge field $V_\mu^a = (A^{-1})_\mu~^\nu B_\nu^a $
which itself transforms as the standard realization and in terms of which the  SUSY and gauge 
covariant derivative takes the form 
\be
({\cal D}_\mu \psi)^i =e^{-\frac{i}{f_a}n_\psi  a }[(D_\mu \psi)^i + T^a_{ij} V^a_\mu \psi^j ] ~.
\ee
In a similar fashion, the gauge covariant field strength can also be brought into the standard realization by defining ${\cal F}_{\mu\nu}^a = (A^{-1})_\mu~^\alpha (A^{-1})_\nu~^\beta F_{\alpha\beta}^a$, where $F_{\alpha\beta}^a = \partial_\alpha B_\beta^a -\partial_\beta B_\alpha^a +if_{abc}B_\alpha^b B_\beta^c $ is the usual field strength having the nonlinear SUSY transformation 
$\delta_Q(\xi,\bar{\xi}) F_{\mu\nu}^a = 
\Lambda^\rho (\xi,\bar\xi)\partial_\rho F_{\mu\nu}^a  +\partial_\mu 
\Lambda^\rho (\xi,\bar\xi)F_{\rho\nu}^a +\partial_\nu \Lambda^\rho (\xi,\bar\xi) F_{\mu\rho}^a $~. 

Using as basic building blocks the gauge singlet Goldstino and R-axion nonlinear SUSY covariant derivatives, the matter fields and their SUSY-gauge covariant derivatives and the standard realization field strength tensor, a nonlinear SUSY, R and gauge invariant action can be constructed as
\bea
\Gamma&=&\int d^4x \,\, (det A) \,\, {\cal O}(\nabla_\mu \lambda, 
\nabla_\mu \bar{\lambda}, D_\nu a, \psi^i, {\cal D}_\mu \psi^i, {\cal F}_{\mu\nu})
\eea
where ${\cal O}$ is any Lorentz and gauge invariant function. Included in ${\cal O} $ is the  Standard Model Lagrangian or its supersymmetric extension modified only by the replacement of all derivatives by nonlinear SUSY covariant derivatives and field strengths by standard realization field strengths. These invariant action terms dictate the couplings of the Goldstino and R-axion which carry the residual consequences of the spontaneously broken supersymmetry and R-symmetry.

For linearly realized SUSY, a considerable amount of the underlying theoretical structure can be gleaned from the symmetry currents. A general Noether current (composed of generic fields $\phi^i$) has the form 
\be
J^\mu =-\sum_i \Delta \phi^i \frac{\partial {\cal L}}{\partial \partial_\mu \phi^i} -\Delta x^\nu T^\mu~_\nu \, .
\ee
where 
\be
T^\mu~_\nu =-\sum_i \partial_\nu \phi^i \frac{\partial {\cal L}}{\partial\partial_\mu \phi^i}+\eta^\mu_\nu {\cal L}
\ee
is the canonical energy-momentum tensor. Here $\eta^{\mu\nu}$ is the Minkowski metric tensor with signature $(-,+,+,+)$.  Application of Noether's theorem then gives the current divergence 
\be
\partial_\mu J^\mu = -\Delta{\cal L}-(\partial_\mu \Delta x^\mu) {\cal L} + \sum_i \delta \phi^i \frac{\delta \Gamma}{\delta \phi^i}\, ,
\ee
where the last term on the RHS is simply the Euler-Lagrange derivative $\frac{\delta \Gamma}{\delta\phi^i}= \frac{\partial {\cal L}}{\partial \phi^i} -\partial_\mu \frac{\partial {\cal L}}{\partial \partial_\mu \phi^i}$ which vanishes on shell. 

For the model described by the Lagrangian of Eq. (\ref{L1}), the conserved canonical energy-momentum tensor\cite{CL1},\cite{CL89} is simply 
\be
T^\mu~_\nu = 
A^{-1}~_\nu~^\mu {\cal L} + (\det{A})~ D^\rho a A^{-1}_\rho~^\mu  D_\nu a
\ee
Note that the energy-momentum tensor starts as a positive vacuum energy, $<0|T^{00}|0> =  \frac{f_s^4}{2} $,  associated with the spontaneous SUSY breaking. Using the Noether construction, it is straightforward to obtain the form of the supersymmetry and $R$ currents along with their (non-) conservation laws. The conserved supersymmetry currents are 
\be
Q^\mu(\xi,\bar\xi) =\xi^\alpha Q^\mu_\alpha +\bar{Q}^\mu_{\dot\alpha}\bar\xi^{\dot\alpha}=2T^\mu~_\nu \Lambda^\nu (\xi,\bar\xi) \, ,
\ee
with $Q^\mu_\alpha =\frac{2i}{f_s^2} T^\mu~_\nu (\sigma^\nu \bar\lambda)_\alpha ~~;~~\bar{Q}^\mu_{\dot\alpha}=-2i T^\mu~_\nu (\lambda\sigma^\nu)_{\dot\alpha}$, while the $R$ current  
\be
R^\mu = f_a (\det{A})~ A^{-1}_\rho~^\mu D^\rho a - \frac{2}{f_s^4} T^\mu~_\nu (\lambda \sigma^\nu \bar\lambda)
\ee
 has (on-shell) divergence
\be
\partial_\mu R^\mu = m_a^2 f_a (\det{A})~ a 
\ee
displaying the soft $R$ symmetry breaking. Note that $R^\mu = f_a \partial^\mu a +...$ and $Q^\mu_\alpha = i f_s^2(\sigma^\mu \bar\lambda)_\alpha +... ~;~\bar{Q}^\mu_{\dot\alpha} = -if_s^2 (\lambda \sigma^\mu )_{\dot\alpha)}+...$ 
so that $R^\mu$ interpolates for the $R$-axion field $a$, while the supersymmetry currents interpolate for the Goldstino fields $\lambda_\alpha, \bar\lambda_{\dot\alpha}$.

For linearly realized supersymmery, the various currents are components of a supercurrent\cite{FZ}  and are related via SUSY transformations. Under nonlinear supersymmetry, the currents transform\cite{CL1},\cite{CL89} as 
\bea
&&\delta_Q(\xi,\bar\xi) R^\mu =i(\xi^\alpha Q^\mu_\alpha  -\bar{Q}^\mu_{\dot\alpha}\bar\xi^{\dot\alpha}) +\partial_\rho [\Lambda^\rho (\xi,\bar\xi) R^\mu -
\Lambda^\mu (\xi,\bar\xi) R^\rho] + \Lambda^\mu (\xi,\bar\xi) \partial_\rho R^\rho \nonumber \\
&&\delta_Q(\xi,\bar\xi)Q^\mu_\alpha = 2i (\sigma^\nu \bar\xi )_\alpha T^\mu~_\nu +\partial_\nu [\Lambda^\nu(\xi,\bar\xi)Q^\mu_\alpha -\Lambda^\mu(\xi,\bar\xi)Q^\nu_\alpha ] +\Lambda^\mu(\xi,\bar\xi)\partial_\nu Q^\nu_\alpha \nonumber \\
&&\delta_Q(\xi,\bar\xi)\bar{Q}^\mu_{\dot\alpha} = -2i (\xi\sigma^\nu )_{\dot\alpha} T^\mu~_\nu +\partial_\nu [\Lambda^\nu(\xi,\bar\xi)\bar{Q}^\mu_{\dot\alpha} -\Lambda^\mu(\xi,\bar\xi)\bar{Q}^\nu_{\dot\alpha} ] +\Lambda^\mu(\xi,\bar\xi)\partial_\nu \bar{Q}^\nu_{\dot\alpha} \nonumber \\
&&\delta_Q(\xi,\bar\xi) T^\mu~_\nu = \partial_\rho [\Lambda^\rho(\xi,\bar\xi)T^\mu~_\nu-\Lambda^\mu(\xi,\bar\xi)T^\rho~_\nu]+\Lambda^\mu(\xi,\bar\xi)\partial_\rho T^\rho~_\nu \, .
\eea
Being algebraically divergenceless, the total derivative terms on the RHS are Belinfante improvements which can absorbed into defining improved currents. Thus, under the nonlinear SUSY, $R^\mu$ transforms into  $Q^\mu_{\alpha}, \bar{Q}^\mu_{\dot\alpha}$ and $Q^\mu_{\alpha}, \bar{Q}^\mu_{\dot\alpha}$ transforms into $T^\mu~_\nu$. This holds even though $a$ and $\lambda, \bar\lambda$ need not be SUSY partners in the underlying theory, the dynamics responsible for SUSY breaking and $R$ breaking may have different origins and $f_s$ need not equal $f_a$. Since they are related by SUSY transformations, there will be relations among  $R$ and SUSY current correlators. Moreover, as these currents 
interpolate for the $R$-axion and Goldstinos, the current correlator relations will translate into relations among Green functions containing $R$-axions and Goldstinos.

For linear realizations of supersymmetry, not only are the $R$ current, supersymmetry currents and the energy momentum tensor related by supersymmetry transformations, but the explicit breakings of the $R$ and dilatation symmetries are also related via SUSY transformations to the breaking of the SUSY conformal symmetry. To examine the analog of this connection for nonlinear SUSY, we need the variations under the full  superconformal algebra\cite{WZ},\cite{F},\cite{CL1} which, in addition to the SUSY algebra generators, contains  the dilatation generator, $D$, the special conformal generators, $K^\mu$ the SUSY conformal generators, $S_\alpha, \bar{S}_{\dot\alpha}$ as well as generators of angular momentum, $M^{\mu\nu}$. The total variations of the Goldstino and $R$-axion fields under these transformations are secured\cite{CL1},\cite{sconformal} as
\bea
&&\Delta_D \lambda^\alpha =-\frac{1}{2}\lambda^\alpha ~;~ \delta_D \bar\lambda^{\dot\alpha} =-\frac{1}{2}\bar\lambda^{\dot\alpha} ~;~ \delta_D a= 0 \cr
&&\Delta_{S\alpha} \lambda^\beta = \frac{4i}{f_s^2}\lambda^\alpha\lambda^\beta~;~ \Delta_{S\alpha}\bar\lambda^{\dot\beta}=-f_s^2 x^\nu\bar\sigma_\nu^{\dot\beta \alpha}-\frac{2i}{f_s^2}\lambda^\alpha\bar\lambda^{\dot\beta}~;~\Delta_{S\alpha} a = \frac{3f_a}{f_s^2}\lambda_\alpha \cr
&&\bar\Delta_{\bar{S}\dot\alpha} \lambda^\beta = f_s^2x^\nu\bar\sigma_\nu^{\dot\alpha \beta} +\frac{2i}{f_s^2}\bar\lambda^{\dot\alpha}\lambda^{\beta}~;~\bar\Delta_{\bar{S}\dot\alpha} \bar\lambda^{\dot\beta}= -\frac{4i}{f_s^2} \bar\lambda^{\dot\alpha}\bar\lambda^{\dot\beta}~;~ \bar\Delta_{\bar{S}\dot\alpha} a = \frac{3f_a}{f_s^2}\bar\lambda^{\dot\alpha} \cr
&&\Delta_{K\mu} \lambda^\alpha = (x_\mu -\frac{2i}{f_s^4}\lambda\sigma_\mu\bar\lambda)\lambda^\alpha -ix^\nu(\sigma_{\mu\nu}\lambda)^\alpha ~;~\Delta_{K\mu} a = -\frac{3f_a}{f_s^4}\lambda\sigma_\mu\bar\lambda  \cr
&&\Delta_{K\mu} \bar\lambda^{\dot\alpha} = (x_\mu -\frac{2i}{f_s^4}\lambda\sigma_\mu\bar\lambda)\bar\lambda^{\dot\alpha} -ix^\nu(\bar\sigma_{\mu\nu}\bar\lambda)^{\dot\alpha} ~,
\label{SCV}
\eea 
while the space-time point moves  as 
\bea
&&\Delta_D x^\mu =-x^\mu~;~\Delta_{K_\nu} x^\mu = -[\eta_\mu^\nu x^2 -2 x^\nu x_\mu +\eta^\nu_\mu\frac{1}{f_s^8}(\lambda\lambda) (\bar\lambda\bar\lambda)]\cr
&&\Delta_{S\alpha} x^\mu = \frac{i}{f_s^2}(\sigma^\nu \bar\sigma^\mu \lambda)_\alpha x_\nu -\frac{2}{f_s^6}\lambda_\alpha (\lambda\sigma^\mu\bar\lambda)\cr
&&\bar\Delta_{\bar{S}\dot\alpha} x^\mu = -\frac{i}{f_s^2} (\bar\lambda \bar\sigma^\mu\sigma^\nu)_{\dot\alpha} x_\nu -\frac{2}{f_s^6}\bar\lambda_{\dot\alpha}(\lambda\sigma^\mu\bar\lambda) ~.
\eea
The intrinsic variations are then readily obtained using Eq. (\ref{IV}) and the above total variations.
Since the dilatation variations form a linear representation, the scale symmetry is not spontaneously broken. The scaling weights are extracted as $d_\lambda = d_{\bar\lambda}=-\frac{1}{2}~~;~~d_a = 0 $. In fact, the Nambu-Goldstone particle associated with any spontaneously broken symmetry which commutes with the dilatation charge is constrained to have scaling weight zero\cite{CLL} as is the case here for the $R$ axion. With the variations in hand, the dilatation current is constructed as 
\be
{\cal D}^\mu = x^\nu T^\mu~_\nu
\ee
 while its (on-shell) divergence 
\bea
\partial_\mu {\cal D}^\mu &=&[f_s\frac{\partial}{\partial f_s}+f_a\frac{\partial}{\partial f_a}+m_a\frac{\partial}{\partial m_a}]{\cal L}
\eea
exhibits the explicit scale symmetry breaking. Note that there are independent breakings arising from the spontaneous SUSY breaking scale, $f_s$, the spontaneous $R$ symmetry breaking scale, $f_a$, and the soft $R$ symmetry breaking mass term, $m_a$. The special conformal transformations, although nonlinear, have no constant terms. Thus the special conformal tranformations are also not spontaneously broken in this realization. 

Since the $\bar{S} (S) $ variations of $\lambda(\bar\lambda)$ start as constants, it follows that the SUSY conformal symmetries are also spontaneously broken. However, the associated Nambu-Goldstone fermions, $\lambda_S^\alpha~,~\bar\lambda_S^{\dot\alpha}$ are not independent dynamical degrees of freedom. Rather, one finds that $\lambda_S^\alpha = -\frac{1}{4} (\sigma^\mu\partial_\mu\bar\lambda)^\alpha +... ~~;~~\bar\lambda_S^{\dot\alpha} = -\frac{1}{4} (\partial_\mu\lambda\sigma^\mu)^{\dot\alpha} +...$ so that $
\Delta_{S\alpha}\lambda_S^\beta = f_s^2\delta_\alpha^\beta +... ~~;~~\bar\Delta_{S\dot\alpha}\bar\lambda_{\bar{S}}^{\dot\beta} = -f_s^2\delta_{\dot\alpha}^{\dot\beta} +... $. The fact that there can be spontaneously broken spacetime symmetries without independent Nambu-Goldstone fields\cite{inversehiggs} is  not in conflict with Goldstone's theorem\cite{G} which guarantees an independent Nambu-Goldstone field for every spontaneously broken global symmetry. Applying Noether's theorem, the superconformal currents are then computed as 
\bea
&&S^\mu(\eta,\bar\eta)=  [(\bar\eta \bar\sigma^\nu)^\alpha Q_{\mu \alpha } +\bar{Q}_{\mu \dot\alpha}(\bar\sigma^\nu\eta)^{\dot\alpha}]x_\nu +(\eta\lambda+\bar\eta\bar\lambda)[\frac{3}{f_s^4}R^\mu +\frac{2}{f_s^6}T^\mu~_\nu (\lambda\sigma^\nu\bar{\lambda})]~.\nonumber
\eea
Under nonlinear SUSY, the divergence of these currents transform as
\bea
&&\delta_Q(\xi,\bar\xi) \partial_\mu S^\mu (\eta,\bar\eta) =3(\eta\xi+\bar\eta\bar\xi)\partial_\mu R^\mu +2i(\eta\xi-\bar\eta\bar\xi)\partial_\mu{\cal D}^\mu \cr
&&~ -(\xi\sigma^{\nu\rho}\eta + \bar\eta \bar\sigma^{\nu\rho}\bar\xi)\partial_\mu M^\mu~_{\nu\rho}  +\partial_\mu[\Lambda^\mu (\xi,\bar\xi)\partial_\nu S^\nu (\eta,\bar\eta)]
\eea
where
\be
M^\mu~_{\nu\rho}= T^\mu~_\nu x_\rho - T^\mu~_\rho x_\nu -\frac{1}{2f_s^4}[\lambda \sigma^\lambda\bar\sigma_{\nu\rho}\bar\lambda +\lambda\sigma_{\nu\rho}\sigma^\lambda\bar\lambda]T^\mu~_\lambda 
\ee
is the conserved angular momentum tensor. Thus the divergence of the $R$-symmetry current and the divergence of the dilatation current is related to the divergence of the superconformal current through a nonlinear SUSY transformation. Since the divergences of these currents give the explicit breakings of these symmetries in the action, these breakings are also related via the nonlinear SUSY transformation. 

Now suppose the dynamics also generates a spontaneously broken internal global chiral symmetry with associated Nambu-Goldstone boson fields $\pi^i$ which transform under nonlinear SUSY as the standard realization.
A nonlinear SUSY invariant action which also contains a soft explicit chiral symmetry breaking mass term is readily constructed as
\bea
&&\Gamma=\int d^4x \{-\frac{f_s^4}{2}(\det{A})~-\frac{f_a^2}{2}(\det{A})~D_\mu a D^\mu a -\frac{1}{2}m_a^2 f_a^2 (\det{A})~ a^2 \cr
&&~~-\frac{F_\pi^2}{4}(\det{A})~{\rm Tr}\left[ D_\mu U^\dagger D^\mu U\right] +\frac{<\bar\psi \psi>}{2} (\det{A})~{\rm Tr}\left[mU^\dagger +Um\right] \}
\nonumber
\eea
where $U(\pi)=e^{2iT^i \pi^i/f_\pi }$ with  $T^i$ the fundamental representations of the broken generators of the internal symmetry group, $f_\pi$ is the Nambu-Goldstone boson decay constant, $m$ is the mass matrix characterizing the soft explicit chiral symmetry breaking and $\psi$ a chiral fermion of the underlying theory whose condensation is responsible for the spontaneous chiral symmetry breaking. 

Finally, consider the case where in addition to the spontaneously broken SUSY and spontaneous and softly broken $R$ symmetry and chiral symmetries, there is a spontaneously broken dilatation symmetry. This results in appearance of the dilaton, $\varphi$, the Nambu-Goldstone boson of spontaneously broken scale symmetry which transforms as a standard realization under the nonlinear SUSY and as a singlet under both $R$ and the chiral symmetry. 
Under dilatations, the combination, $\Sigma(\varphi)=e^{\frac{\varphi}{f_D}}$ transforms as $\Delta_D (\epsilon)\Sigma =\epsilon\Sigma $ which leads to an inhomogeneous scale transformation of the dilaton given by $\Delta_D (\epsilon)\varphi=\epsilon f_D $. The prescription for nonlinearly realizing the scale symmetry is then to introduce
a factor of $\Sigma(\varphi)$ for every dimensionful parameter appearing in the action. This procedure makes the scaling weight of each term in the action equal to four. Note that the Akulov-Volkov vierbein has scaling weight zero. So doing, the leading terms in an effective Lagrangian which nonlinearly realizes SUSY, R-symmetry, chiral symmetry and dilatations takes the form\cite{CLD}
\bea
&&{\cal L} = -\frac{f_s^4}{2} ~(\det{A}) ~\Sigma^4 -\frac{f_a^2}{2} ~(\det{A}) ~\Sigma^2 {D}_\mu a {D}^\mu a \cr
&&~-\frac{1}{4}f_\pi^2 ~(\det{A})~ \Sigma^2 Tr \left[{D}_\mu U(\pi){D}^\mu U^\dagger(\pi)\right] -\frac{1}{2}f_D^2~ (\det{A})~ {D}_\mu \Sigma {D}_\mu \Sigma \nonumber \\
&&~ +\frac{<\bar\psi \psi>}{2} ~(\det{A})~\Sigma^{3-\gamma}{\rm Tr}\left[mU^\dagger +Um\right] -\frac{f_a^2}{2}m_a^2 ~(\det{A})~ \Sigma^2 a^2 ~.\nonumber
\eea
Here the last line includes the soft explicit R and chiral  breakings while $\gamma$ is the anamolous dimension of the underlying fermion field.

Note, however, that since $\det{A}~= 1+...$, this Lagrangian contains an $\Sigma^4$ potential term for the Nambu-Goldstone dilaton which in turn produces a destabilizing linear in $\varphi$ term unless $f_s^4 = \frac{(3-\gamma)}{2} <\bar\psi \psi> Tr[m]$  in which case it is cancelled against the linear in $\varphi$ term arising from the soft expicit chiral symmetry breaking. But then the Goldstino decay constant, $f_s$, vanishes in chiral limit. Since the Goldstino decay constant is the matrix element of the supersymmetry current between the vacuum and the Goldstino state, its vanishing signals the nonviability of the Goldstone realization. Consequently, scale symmetry and SUSY symmetries cannot be simultaneously nonlinearly realized and the spectrum cannot contain both a dilaton and a Goldstino as Nambu-Goldstone particles. Note that there is no difficulty in constructing an effective Lagrangian invariant under both nonlinear SUSY and chiral symmetry transformations nor under nonlinearly realized scale symmetry and chiral symmetry\cite{WB}. It is only when both SUSY and scale symmetry are to be realized nonlinearly that one encounters the inconsistency.

$\hspace{.2in}$

\noindent
This work was supported in part by the U.S. Department of Energy 
under grant DE-FG02-91ER40681A29 (Task B). I thank T. Clark for many enjoyable discussions.

\end{document}